%% file: 0_main.tex
\documentclass[letterpaper,10pt]{article}
\usepackage[top=1in,
      bottom=1in,
      left=1in,
      right=1in]{geometry}

\usepackage{neutral_style}
\usepackage[kerning,spacing]{microtype}
   \microtypecontext{spacing=nonfrench}

\usepackage[T1]{fontenc}

\usepackage{graphicx} 
\usepackage[table,svgnames,x11names]{xcolor}

\usepackage{url}
\usepackage{hyperref}
  \hypersetup{colorlinks=true, allcolors=DarkBlue}

\usepackage{listings}
\usepackage{tikz}
  \usetikzlibrary{shadows}
  \usetikzlibrary{shadows.blur}

\usepackage{tcolorbox}
  \tcbuselibrary{breakable,listings,skins}        
  \newtcblisting{prompt}{listing only,skin=enhanced,width=0.95\linewidth,boxrule=0.5pt,
                         colback=black!5!white,colframe=black!40!white,
                         before=\begin{center},after=\end{center},boxsep=0mm,
                         listing options={style=tcblatex,breaklines=true,breakindent=0pt,},
                         }

\usepackage{pgfplotstable}
  \pgfplotsset{compat=1.18}
\usepackage{booktabs}
\usepackage{colortbl}
\usepackage{tabularx}
\usepackage{multirow}
\usepackage{amssymb}

\usepackage{todonotes}
   \setuptodonotes{size=\footnotesize}

\usepackage{enumitem}
  \newlist{todolist}{itemize}{2}
  \setlist[todolist]{label=$\square$}

\newtcolorbox{violationbox}{
  enhanced, breakable,
  boxrule=0.6pt,
  colback=black!3,
  colframe=red!60,
  arc=0pt,
  left=6pt,right=6pt,top=6pt,bottom=6pt
}

\newtcolorbox{defensebox}{
  enhanced, breakable,
  boxrule=0.6pt,
  colback=black!3,
  colframe=black!60,
  arc=0pt,
  left=6pt,right=6pt,top=6pt,bottom=6pt
}

\usepackage{multicol}

\usepackage{adjustbox}
\usepackage{array}

\usepackage{cuted}
\usepackage{caption}

\usepackage{booktabs}
\usepackage{rotating}
\usepackage{graphicx}
\usepackage{pgfplotstable}

\microtypecontext{spacing=nonfrench}

\usepackage[section]{placeins}

\usepackage{pifont}
\usepackage{inconsolata}

\title{Systems Security Foundations for Agentic Computing}

\author{
Mihai Christodorescu$^{1}$ \quad
Earlence Fernandes$^{2}$ \quad
Ashish Hooda$^{1}$ \quad
Somesh Jha$^{1,3}$ \\
Johann Rehberger$^{4}$ \quad
Kamalika Chaudhuri$^{5}$ \quad
Xiaohan Fu$^{2,6}$ \quad
Khawaja Shams$^{1}$ \\
Guy Amir$^{7}$ \quad
Jihye Choi$^{3}$ \quad
Sarthak Choudhary$^{3}$ \quad
Nils Palumbo$^{3}$ \\
Andrey Labunets$^{2}$ \quad
Nishit V. Pandya$^{2}$ \\
\\
\small
$^{1}$Google \quad
$^{2}$University of California San Diego \quad
$^{3}$University of Wisconsin--Madison \\
\small
$^{4}$EmbraceTheRed \quad
$^{5}$FAIR at Meta \quad
$^{6}$Gray Swan AI \quad
$^{7}$Cornell University
}

\date{}

\begin{document}

\maketitle

\begin{abstract}
\end{abstract}

\input{1_summary}
\input{2_intro}
\input{4_challenges}

\input{5_securitycasestudies}

\input{6_current_approaches}

\input{7_open}
\input{8_conclusions}

\bibliographystyle{plainurl}
\bibliography{references}

\appendix
\input{11_appendix_A}

\input{12_knownsecurity}

\end{document}

%% file: 1_summary.tex
In recent years, \textit{agentic artificial intelligence (AI)} systems are becoming increasingly widespread. 

These systems allow agents to use various \textit{tools}, such as web browsers, compilers, and more.
However, despite their popularity, agentic AI systems also introduce a myriad of security concerns, due to their constant interaction with third-party servers. For example, a malicious adversary can cause  data exfiltration by executing prompt injection attacks, as well as other unwarranted behavior.
These security concerns have recently motivated researchers to improve the safety and reliability of agentic systems. However, most of the literature on this topic is from the AI standpoint and lacks the system-security perspective and guarantees.
In this work, we begin bridging this gap and present an analysis through the lens of classic cybersecurity research.
Specifically, 
motivated by decades of  progress in this domain, we identify short- and long-term research problems in agentic AI safety by examining end-to-end security properties of \textit{entire systems}, rather than standalone AI models running in isolation.
Our key goal is to examine where research challenges arise when applying traditional security principles in the context of AI agents and, as a secondary goal, distill these ideas for AI practitioners.
Furthermore, we extensively cover 11 case studies of real-world attacks on agentic systems, as well as define a series of new research problems that are specific to this important domain.

%% file: 2_intro.tex
\section{Introduction}

Agentic AI systems have become widespread: by being given access to tools such as web browsers, compilers, and more, AI agents, and especially those based on large language models (LLM), are becoming increasingly powerful.
However, this comes at a cost: access to third party tools and external servers introduces a wide range of security challenges that must be addressed in order to safely and reliably deploy such systems in a trustworthy manner.
Previous literature has raised the important topic of agentic AI security, however, most of the this literature is written from an AI perspective, and does not leverage the extensive research accumulating through dozens of years of classic security studies.
In this Systematization of Knowledge (SoK) paper, we begin bridging this gap, and explore how to build upon well-studied concepts and practices in computer-security research to advance the security of agentic computing. We focus on key similarities and differences between both domains, and distill lessons, open research problems and recommendations on how the two fields can co-operate to build a foundation for secure agentic computing.  

\textit{Agentic computing} refers to AI models that interact with computer systems and services on behalf of human users by repeatedly calling tools in a loop~\cite{willison2025agentdefinition}. 
The user specifies a task in natural language, and the AI model (typically a large language or vision model) completes it by iteratively invoking tools, such as desktop UIs, web browsers, or mobile operating systems. 
For instance, a user might ask an agent to \textit{find mentions of a particular event in recent emails, summarize what happened, and save the summary to a text file}. Such a request can trigger tools calls including email search, retrieval, and writes to local storage.
Recently, there is also a trend in promoting uniform standards and protocols for defining various agentic tool calls, e.g., REST or Model Context Protocol (MCP)~\cite{modelcontextprotocol_intro2025}. 

Current efforts to secure agentic systems span multiple layers of defense~\cite{beurerkellner2025designpatternssecuringllm}, including model-level, system-level, and user-level approaches.
At the \emph{model level}, various methods aim to make the agent more resilient to attacks by hardening approaches, often described as alignment techniques~\cite{chen2025secalign,wallace2024instruction,openai2025safetyalignment}.
At the \emph{system level}, an increasingly prominent line of work treats the model as an untrusted (or at best, probabilistic) component and instead enforces invariants through isolation and mediation: restricting the agent's action space, sandboxing tool execution, structuring tool interfaces, and separating untrusted data from control flow via explicit trust boundaries and guardrails~\cite{beurerkellner2025designpatternssecuringllm,zhang2025llmagentsemploysecurity}.
Finally, at the \emph{user level}, practical deployments commonly incorporate governance mechanisms such as explicit approvals/confirmations for sensitive actions, permission scoping, and audit/monitoring to reduce the blast radius of failures and support incident response~\cite{beurerkellner2025designpatternssecuringllm}.

From a computer-security perspective, this emerging taxonomy mirrors a classic \emph{defense-in-depth} lesson: hardening a component is valuable, but is rarely sufficient on its own. Robust systems rely on \emph{deterministic} enforcement points (e.g., access control, isolation, and provenance boundaries) placed at multiple layers of abstraction (hardware, hypervisors, operating systems, user-space frameworks, applications, and networks), to increase the attacker’s burden while preserving performance and functionality~\cite{saltzer,ross-anderson}. In agentic settings, however, deciding \emph{where} to place such enforcement points is less straightforward, because the model simultaneously (i) processes untrusted content and (ii) synthesizes actions and tool calls on the fly.

This tension is aggravated further due to modern agents being vulnerable to inference-time attacks that can steer behavior in arbitrary ways~\cite{zou2023universaltransferableadversarialattacks,fu2024impromptertrickingllmagents,szegedy2014intriguingpropertiesneuralnetworks,greshake2023}. As with earlier waves of adversarial ML, purely behavioral defenses that rely on the model to reliably identify and ignore malicious instructions can remain brittle under adaptive attackers~\cite{nasr2025attackermovessecondstronger,athalye2018obfuscatedgradientsfalsesense,pandya2025iattentionbreakingfinetuning}. Moreover, in the LLM era, attackers often need not have deep system expertise: simple prompting can be enough to induce harmful tool use or policy violations in real-world deployments~\cite{greshake2023,wunderwuzzi2024breaking,wunderwuzzi2025claude,wunderwuzzi2024chatgpt,wunderwuzzi2025devin}. Therefore, to make foundational and practical progress, the fields of computer security and AI agent security need to co-operate: combining model-level robustness with principled system-level and user-level guardrails grounded in decades of security research.

\medskip
\noindent\textbf{Summary of Findings.}
Computer security makes progress by articulating clear security invariants and then enforcing them via layered, largely deterministic mechanisms and design principles (e.g., least privilege, complete mediation, separation of privilege)~\cite{saltzer,ross-anderson}. 
When we attempt to apply these ideas to agentic systems, we encounter several challenges that arise. 
We use these frictions to organize the report and to surface concrete research questions.
Furthermore, each of these challenges also poses interesting research avenues for long-term future work.

\smallskip
\noindent\textit{\underline{(1) Probabilistic and Opaque TCB.}}
A classic systems-security move is to minimize a \emph{trusted computing base (TCB)} that deterministically enforces an invariant (e.g., preventing execution from non-executable memory regions to reduce the impact of memory-corruption bugs). 
In many \emph{AI-based} systems, including agentic ones, the model itself becomes a central part of the TCB, yet its behavior is fundamentally probabilistic and often not fully interpretable. 
If we ``\textit{train the model to never do $X$}'', that guarantee is inherently conditional and statistical: it holds with some probability that depends on the input distribution and the surrounding context rather than as an absolute property of the system. 
The resulting challenge is to build defenses with meaningful assurance guarantees despite relying on a probabilistic/opaque component in the TCB.

\smallskip
\noindent\textit{\underline{(2) Dynamic Security Policies.}}
A security policy specifies what an untrusted program is allowed to do and is typically authored by a developer or inferred from code analysis. 
However, in agentic settings, both these assumptions break down: (i) there may be no developer intent beyond the end-user's natural-language request, and (ii) there is no fixed program to analyze, since the agent effectively synthesizes many programs on the fly. 
Inferring or synthesizing a policy from a task description is difficult because tasks are often underspecified, and using the agent's full context window for precise policy inference is nontrivial due to untrusted external content.

\smallskip
\noindent\textit{\underline{(3) Fuzzy Security Boundary.}}
Deterministic guardrails are most effective when enforced at an interface that exposes the right semantic information (e.g., Android permissions at the boundary between apps and protected services). 
In many agentic stacks, however, the boundary between ``decision'' and ``action'' is blurry: the model may directly emit tool calls (sometimes via low-level UI actions) without stable abstraction layers that cleanly separate intent from mechanism. 
As a result, policies written at too low a level (clicks/keystrokes) become brittle, while policies written at too high a level may lack enforceable grounding. 
This can also be viewed as a manifestation of the classic \emph{semantic gap} problem~\cite{bhushan-semantic-gap,brendan-semantic-gap,sommer-paxson}.

\smallskip
\noindent\textit{\underline{(4) Dynamic Instruction Following.}} Prompt injections are a key security problem to solve in agentic systems, but the underlying mechanism of dynamically updating agentic instructions is oftentimes necessary for functionality and thus not necessarily malicious. For example, tool documentation does affect an agent's behavior (by definition because the agent has to learn how to use the tool!). The challenge here is defining when prompt injections are harmful~\cite{agent-dojo} and when they are useful and then distinguishing the two. The analog in computer security is dynamic code loading--a very challenging problem that remains difficult to address today.

\medskip
\noindent\textbf{Scope and organization of this report.} There has been a range of recent and contemporaneous writings on the general theme of applying computer security ideas to agentic computing~\cite{beurerkellner2025designpatternssecuringllm,zhang2025llmagentsemploysecurity}. The key distinction of this report is that we take a deep-dive on the technical challenges that arise when applying concrete security principles and articulate open research problems based on those technical challenges. 
The paper starts out with a discussion of security principles applied to agentic settings and the challenges that arise forthwith (\autoref{sec:challenges}). \autoref{sec:case-studies} presents 11 case studies of real attacks on agentic systems with three goals in mind: (1) to showcase how real attacks are carried out by adversaries without the need for sophisticated technical tools to execute them; (2) to highlight the corresponding security principle(s) that were violated; and (3) to outline concrete security mechanisms needed to stop each such attack. 
We discuss the current practices for mitigating these threats in \autoref{sec:current-approaches} and summarize the open research problems in \autoref{sec:open}.

%% file: 4_challenges.tex
\section{Challenges in Applying Security Engineering to Agentic Computing}
\label{sec:challenges}

\input{tcb_probabilistic}

\input{policy_selection}

\input{security_boundary}

\input{prompt_injection}

%% file: tcb_probabilistic.tex
\subsection{The TCB is Probabilistic}
\label{sec:probabilistic-tcb}

The {\it Trusted Computing Base (TCB)} is a component or assumption that cannot be influenced by the attacker. TCB  is an invariant which is basis of security, such as the no-execute or NX bit in hardware (e.g. data put in region with NX bit set can never be executed or interpreted as instruction). 

In traditional systems, the TCB consists of deterministic components such as hardware, OS kernels, reference monitors that behave predictably. This determinism enables security guarantees: W$\oplus$X memory protection consistently prevents execution of writable memory, and Content Security Policy deterministically blocks unauthorized scripts. 

In contrast, the TCB in agentic systems is built on LLMs that do not guarantee predictable, or even deterministic behavior---the same prompt can yield different outputs due to sampling randomness. Thus a fundamental challenge in realizing this traditional security principle in agentic systems is the TCB being probabilistic or non-deterministic (e.g. imagine whether we could build a memory safety defense on top of a probabilistic NX bit). This probabilistic nature undermines traditional security principles in three ways. First, a probabilistic reference monitor might correctly deny unauthorized access 99\% of the time, but the remaining 1\% can be exploited. Second, unlike traditional systems where security properties can be formally verified, we cannot prove an LLM will always enforce policies correctly. Third, LLMs operate in continuous representation spaces, making them vulnerable to adversarial examples. Unlike traditional systems that have discrete instruction sets with clear boundaries, attackers can use various search and optimization techniques to efficiently find inputs that appear benign but cause policy violations.

To build secure systems on probabilistic LLMs, we must leverage external deterministic information to disambiguate LLM decisions. An example of a system takes this approach  is SkillFence~\cite{10.1145/3517232}. It uses deterministic signals from the user's browsing history and installed apps to correct any mistakes made by a voice assistant's language understanding. However, we believe that building provable defenses on probabilistic and non-deterministic TCB is a challenging---perhaps impossible---task.

%% file: policy_selection.tex
\subsection{The Security Policy Is Dynamic and Task-Specific}
\label{sec:dynamic-policy}

In traditional computer security, a security policy governs the privilege of a piece of code. For example, in Android, the OS developers provide a set of permissions that an app developer can select from. The key aspect is that the app is generally single-purpose, built to perform a fixed set of tasks, and thus, the app developer can create a security policy and present it to the user at the time of installation. Furthermore, one can analyze the app's code to determine why certain permissions are needed and whether they are necessary for the app's stated functionality (thus allowing for analyzing compliance with the principle of Least Privilege).

In agentic systems, there is no app developer and there is no app. Rather, there is only a natural language task specification that can change over time. Thus, the privilege of the agent is dynamic and needs to be predicted from the task description in a secure manner. For example, consider a simple developer task assigned to a browser agent:
\begin{prompt}
comment on GitLab issue 745 that we're done
\end{prompt}
From this task description, the security policy would need to incorporate the following components:
\begin{enumerate}
    \item The agent can navigate to GitLab issue 745;
    \item The agent should have write access to the comment box;
    \item The agent should not have access to anything else in GitLab or the user's session in the browser.
\end{enumerate}
This least-privilege security policy is task-specific and only known once the task is specified.

Continuing this example, let's say that the user specifies a follow-up task for the agent:
\begin{prompt} summarize issue 745 for me \end{prompt}
Now, how should the privilege of the agent change to solve the new task description? One option is to reset the agent's context and redo the policy prediction. This approach is secure but can result in lost utility because the agent's context is erased. Thus, how can we evolve the security policy in a secure manner while there is untrusted data in the context?
An additional problem for AI agents is that privacy consent is fundamentally at odds with agency ~\cite{tiwari_whittaker_2025_ai_agent_ai_spy}, as finding and enforcing the right policy for an AI agent each time requires resolving the trade-off between providing more access for better task performance and restricting access as per least-privilege principle. We believe that the key lies in creating policy languages that are amenable to formalized reasoning so that when new policies are predicted and added to the existing set, we can reason about how the new policies change the overall privilege level of the agent.

Extrapolating from this example, we posit the following research challenges: (1) \textit{domain-specific policy languages} for different types of agents that are amenable to formal analysis and reasoning; (2) \textit{dynamic policy prediction} in these DSLs given natural language tasks, from trusted and untrusted context.

%% file: security_boundary.tex
\subsection{The Security Boundary Is Fuzzy}
\label{sec:fuzzy-boundary}

Identifying the right abstraction at which security policies are enforced is crucial to upholding the complete mediation design principle. In traditional computer systems, we have layered architectures (Hardware---Operating System (OS) Kernel---Process---Network) that offer different levels of abstraction for enforcing security policies. For example, the process-kernel interface allows for SELinux-style MAC policies and the App-Runtime interface allows for higher-level permissions that we see in systems like Android. Agentic systems lack these layers. Rather, they directly use tools given a natural language prompt. There are no layers in between. Thus, enforcing a security policy at the final layer of tool-calls can lead to incomplete or ineffective mediation (i.e., there can be gaps). The key research challenge in applying the complete mediation principle is identifying or creating the right abstraction at which a security policy can be enforced.

We discuss two recent efforts that have started investigating these challenges along two dimensions: (1) creating an abstraction over which security can be enforced; (2) finding the right level of abstraction in an existing agent design.

\textit{CaMeL}~\cite{camel} and \textit{FIDES}~\cite{fides} introduce an agent design that exemplifies the ``solution as code'' approach. To solve a task, the agent first generates code that, when executed, would solve the task specified in natural language. Notice that this introduces a new layer of abstraction: the code needed to solve the task. At that point, one can analyze the code using standard security analysis methods (control and data flows) to determine whether the agent's plan is aligned with the user's goal. A key assumption in this design is that the agent has access to a set of semantically-meaningful tools. One of the key future research questions here is creating an appropriate intermediate representation for agent plans that is amenable to static and dynamic analysis.

\textit{ceLLMate}~\cite{cellmate} introduces a sandboxing framework for browser agents. The key challenge is identifying the right abstraction to enforce security policies within existing infrastructure. Browser agents, as currently built, have access to low-level UI manipulation tools (e.g., clicks and keystrokes), and thus, it is non-sensical to write policies at that level. The key insight to address this semantic gap is realizing that no matter what UI manipulations the agents perform, HTTP-level messages capture those actions in semantically meaningful ways. Therefore, ceLLMate policies are specified and enforced at the HTTP level, in co-operation with information supplied by webserver developers. One of the key future research challenges here is automatically predicting policies at the level of HTTP messages using natural language task descriptions as input.

%% file: prompt_injection.tex
\subsection[Dynamic Instruction Following]{Dynamic Instruction Following}
\label{sec:prompt-injection}

Prompt injection, more specifically indirect prompt injection, is often seen as a vulnerability of current AI-based systems. But more generally, prompt injection can be seen as a manifestation of a new class of functionality in AI, where the agent can adjust its task based on \textit{new instructions} in its context. For example, using the GitLab task in \autoref{sec:dynamic-policy}, the agent may start with the aforementioned summarization request prompt, subsequently visit the GitLab page for that issue, identify that it depends on another issue, and decide to summarize both to give the user a comprehensive answer. The distinction between (undesirable) prompt injection and (desirable) contextual task adjustment rests on the provenance of the input that triggered the change and the scope of the change.

From the security perspective, dynamic instruction following, which can be realized as prompt injection (benign or malicious), has a parallel in non-AI systems in the form of dynamic code loading. This is common among applications that support plugins or extensions (e.g., OS kernels, IDEs, web browsers) and has long posed significant challenges in terms of security. The typical solution is to introduce another security boundary between the main TCB and the dynamically loaded code and to use permissions or sandboxing to limit the access afforded to the new code. For example, web pages (probably the most popular platform for dynamic code loading, as each page load and subsequent interaction can potentially load additional Javascript code as web-page scripts) are secured through a number of sandboxes and associated security policies: Content Security Policy determines which third-party scripts can be loaded, Same-Origin Policy restricts the web page data accessible to a third-party script, \texttt{<iframe>} sandboxing isolates third-party content, and Subresource Integrity ensures that third-party scripts have not been modified from the time it was approved by the developer. Other systems, such as Android, actively discourage dynamic code loading~\cite{android_dcl}.

In the agentic space, dynamic instruction following is a key feature to steer task execution. For example, MCP descriptions are instructions brought in externally, as are Anthropic's Claude Skills~\cite{claude_skills} and OpenClaw's ClawHub add-ons~\cite{clawhub}. Yet current agentic systems lack both of the components that provide security in web page dynamic code loading. First, information about the source (or provenance) of an instruction given to the agent is hard to assess (in contrast to the source of web-page scripts, most commonly loaded through explicit URLs over HTTPS). Second, sandboxing in the agent context is unavailable or probabilistic at best via mechanisms such as Instruction Hierarchy~\cite{wallace2024instruction}. An additional challenge to further drive the contrast with the web-page security model is that web pages are designed by a website owner, who is in charge of determining which code to include on the page and indirectly which code to dynamically load at some later time on the page. Agents are often given underspecified instructions and are expected to refine them with additional instructions discovered while working on the task at hand. This is a highly desirable functionality of agentic systems (not having to fully specify the task completely in the first prompt), but one for which security mechanisms are not readily available.

%% file: 5_securitycasestudies.tex
\section{Attacks on Agentic Systems: Case Studies}
\label{sec:case-studies}

We present here 11 attacks against agentic systems to illustrate the variety of types of vulnerabilities that occur in practice. For each attack we also highlight the security principle(s) that were violated and the potential defenses. For a refresher on the security principles we reference here, please see \autoref{sec:known}. A summary of these attacks and corresponding violated security principles is shown in \autoref{tab:attacks_principles}.

\paragraph{Microsoft Copilot Exfiltration.} A vulnerability in Microsoft 365 Copilot allowed attackers to steal private user data, such as emails, by sending a malicious message containing a hidden prompt~\cite{wuest2024m365}. When a user interacted with this message (asking, for example, for a summary) via Copilot, a prompt-injection attack was triggered, letting the attacker take control of the agent. The compromised Copilot was then instructed to find sensitive information, encode it using ``ASCII smuggling,'' and embed it into a seemingly harmless hyperlink. When the user clicked this link, their data was secretly sent to the attacker.
\noindent
\begin{violationbox}
\underline{Violation:}
This attack violates the security principles of \textit{Least Privilege}, \textit{Complete Mediation}, and \textit{Secure Information Flow}, as Copilot automatically performed unexpected actions sourced from a document of unknown origin (like searching for and exfiltrating data) without verifying each step with the user, failing to check that the AI's operations were fully authorized. This vulnerability is similar to that of traditional code injection after a buffer overflow, where an adversarially crafted input allows the attacker to run code of their choice inside the victim process, coupled with insufficient access control.
\end{violationbox}

\begin{defensebox}
\underline{Defense:} 
Implement strict output sanitization to detect and block data ex-filtration channels like ``ASCII smuggling'' in generated hyperlinks. Require human approval whenever the agent accesses sensitive data. Because the TCB is probabilistic (the CoPilot LLM is part of the TCB, \autoref{sec:probabilistic-tcb}) and the security boundary for Internet access is fuzzy (the set of safe URLs cannot be practically enumerated in full, \autoref{sec:fuzzy-boundary}), these defenses provide only incomplete security guarantees and will need to be supplemented with mechanisms for separating instructions and data (\autoref{sec:open-separate}) and for least-privilege access control (\autoref{sec:open-plop}).
\end{defensebox}

\input{tables/attacks_principles.tex}

\paragraph{Devin AI Exposed Ports.}
The AI agent Devin comes with tool called \lstinline!expose_port!, meant for testing, that was abused through an indirect prompt injection attack~\cite{wunderwuzzi2025devin}. An attacker hosted a malicious prompt on a website that, when visited by Devin, hijacked the agent. The compromised AI then started a local web server, exposing its entire file system, and used the \lstinline!expose_port! tool to make this server publicly accessible online. The resulting URL was then sent to the attacker, granting them full access to Devin's files.
\noindent
\begin{violationbox}
\underline{Violation:}
This vulnerability violates the principles of Least Privilege (as the \lstinline!expose_port! tool had excessive permissions, allowing it to expose any port, including one with access to the entire file system, rather than being restricted to only what was necessary for its intended function) and of Security Information Flow (as the agent accepted instructions received from an unknown origin). This vulnerability is similar to that of traditional code injection after a buffer overflow.
\end{violationbox}

\begin{defensebox}
\underline{Defense:} 
A potential solution is to restrict the \lstinline!expose_port! tool to a predefined safe range of ports by default and configurable only from outside the agent's sandbox. However, such a safe list might be prompt specific and thus presents the challenge of a dynamic, task-specific security policy (\autoref{sec:dynamic-policy}). A strong defense will need controls that guarantee minimum-privilege access, an open problem for agentic systems (see \autoref{sec:open-plop}).
\end{defensebox}

\paragraph{ChatGPT Long-Term Memory SpAIware.}
A vulnerability in the ChatGPT macOS app allowed for persistent data exfiltration by injecting malicious instructions into the app's ``Memories'' feature~\cite{wunderwuzzi2024chatgpt}. This was accomplished through a prompt injection in an untrusted document or website, which caused the application to continuously send all of the user's conversations to an attacker-controlled server. The data was exfiltrated by rendering an invisible image that included the user's data as a parameter in the image URL.
\noindent
\begin{violationbox}
\underline{Violation:}
This attack violates the principles of \textit{Trusted Computing Base (TCB) Tamper Resistance} and Secure Information Flow because it stores data from an untrusted origin into the Memories storage (presumably trusted by the agent) and creates an unauthorized channel for sensitive information to be leaked from the application to an external, malicious server. In addition to the similarity with code injection after buffer overflow, this vulnerability also allows the attacker to achieve persistent control, which is one of the fundamental issues with dynamic code loading (\autoref{sec:prompt-injection}).
\end{violationbox}

\begin{defensebox}
\underline{Defense:} 
All data entering ``Memories'' should be sanitized before storage or, at a minimum, passed through a human-in-the-loop check (as other agents do by \textit{suggesting} memories to the user instead of automatically persisting them). Additionally, the app's rendering layer should prevent automatic loading of remote sources from untrusted origins. The deployed fix performs an additional check via a \lstinline!url_safe! tool to control which URLs ChatGPT will connect to, in turn making exfiltration more difficult.. We note that sanitization for untyped data (such as natural-language text) and determining the trust level of a remote Internet resource (such as a web page) both often rely on classifiers, raising concerns of building on a probabilistic TCB (\autoref{sec:probabilistic-tcb}).
\end{defensebox}

\begin{figure*}[t]
    \centering
    \includegraphics[trim={500px 150px 0 0},clip,width=0.85\textwidth]{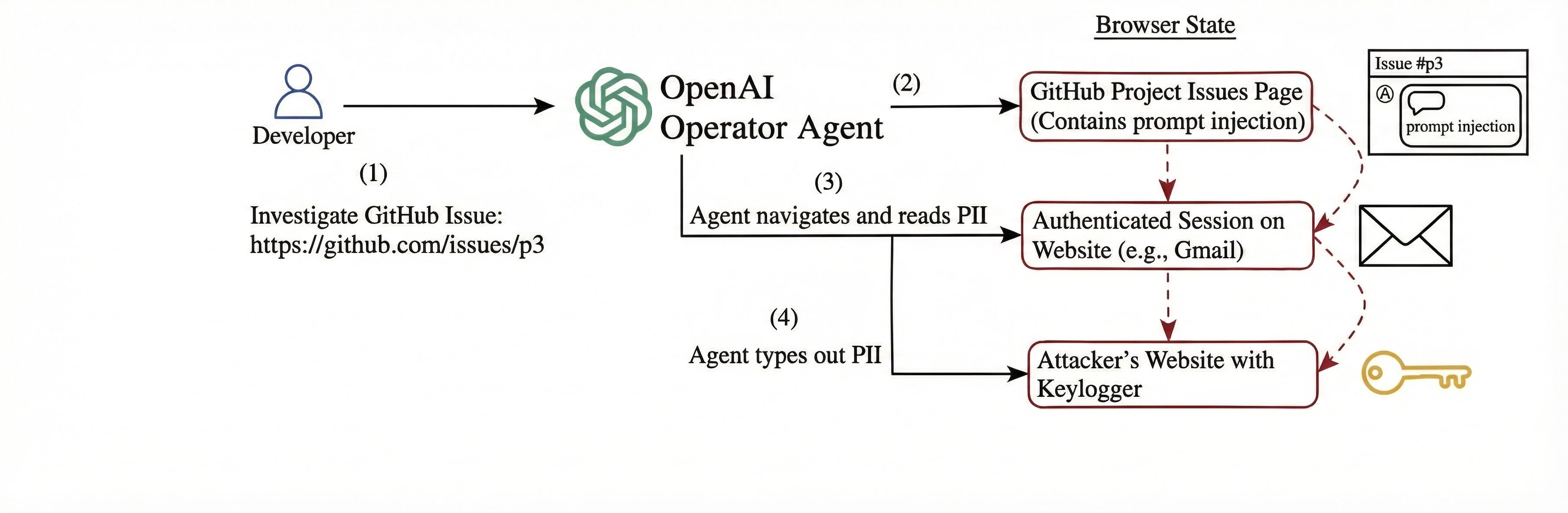}
    \caption{OpenAI Operator exploit flow: a prompt-injected GitHub issues page can steer the agent into an authenticated session (e.g., Gmail) and onward to an attacker-controlled site, resulting in personally identifiable information (PII) exfiltration.}
    \label{fig:operator-flow}
\end{figure*}

\paragraph{Amp~AI Code Arbitrary Command Execution.}
Sourcegraph's Amp AI coding agent allowed for arbitrary command execution by exploiting the agent's ability to modify its own configuration file~\cite{wunderwuzzi2025amp}. Through a prompt injection attack, an adversary instructed the AI to alter its \lstinline!settings.json! file, either by adding malicious commands to an allowlist for automatic execution or by adding an attacker-controlled server to the configuration, both of which led to running unauthorized code on the developer's machine.
\noindent
\begin{violationbox}
\underline{Violation:}
This attack fundamentally violates the principles of TCB Tamper Resistance (as the AI agent, a component of the trusted system, was able to modify its own security-critical configuration files, thereby compromising the integrity of the system's security policies) and Secure Information Flow. In addition to the similarity with code injection after buffer overflow, this vulnerability has parallels with privilege escalation, where the attacker can modify the security configuration of their environment to gain higher access.
\end{violationbox}

\begin{defensebox}
\underline{Defense:} 
Such systems should enforce immutability on security-critical configuration files (like \lstinline!settings.json!) so the agent cannot modify its own execution environment. Any changes to these configurations should require human approval. These defenses highlight the challenges of fuzzy security boundaries (\autoref{sec:fuzzy-boundary}) and dynamic code loading (\autoref{sec:prompt-injection}), for which agentic systems have no guaranteed solutions (see Sections~\ref{sec:open-plop} and~\ref{sec:open-ifc} for further discussion on these open problems).
\end{defensebox}

\paragraph{DeepSeek AI Account Takeover.}
A vulnerability in the DeepSeek AI platform led to a full account takeover by chaining a prompt injection with a Cross-Site Scripting (XSS) exploit~\cite{johann2024deepseek}. An attacker uploaded a malicious text file containing a base64 encoded JavaScript payload, which, when processed by a victim's account, instructed the AI to decode and execute the script in the victim's browser, stealing the \lstinline!userToken! from \lstinline!localStorage! and sending it to the attacker.
\noindent
\begin{violationbox}
\underline{Violation:}
This attack violates the principles of Secure Information Flow (as it established an unauthorized covert channel to leak a sensitive session token from the user's browser to an external, malicious endpoint, completely bypassing the platform's intended data handling and security boundaries) and \textit{Complete Mediation} (as the agent processed encoded data meant to bypass input filtering). Chained code injections are often used in traditional exploits to complete an attack.
\end{violationbox}

\begin{defensebox} 
\underline{Defense:} 
Ensure a strict separation between data (uploaded text files) and executable web code. The fuzzy security boundary (\autoref{sec:fuzzy-boundary}) coupled with the need to allow for some flexibility in using external data as prompts (a form of dynamic code loading, \autoref{sec:prompt-injection}) makes it challenging to secure the agentic system end-to-end when no good mechanisms for separating instruction and data are available (see \autoref{sec:open-separate}).
\end{defensebox}

\paragraph{Terminal DiLLMa.}
The ``Terminal DiLLMa'' attack hijacked a user's terminal through LLM-powered command-line tools by using prompt injection to generate malicious ANSI escape sequences~\cite{wunderwuzzi2024terminal}. When a compromised tool, such as the proof-of-concept \lstinline!dillma.py!, processed a malicious prompt, it output specially crafted ANSI codes that the terminal emulator executes, leading to unauthorized actions like clipboard manipulation or data exfiltration via DNS requests.
\noindent
\begin{violationbox}
\underline{Violation:}
This attack highlights a violation of the Complete Mediation and Secure Information Flow principles. Insufficient or incorrect input validation and sanitization result in incompletely validated inputs which then trigger unwanted software behaviors, as occurred here.
\end{violationbox}

\begin{defensebox}
\underline{Defense:} 
Implement a sanitization layer for agent output for dangerous ANSI escape sequences before they reach the terminal emulator. This becomes challenging as highlighted in \autoref{sec:fuzzy-boundary} when different types of outputs can be created (e.g., with ANSI escape codes, with Markdown formatting, with HTML tag and Javascript) and can be displayed in various settings (e.g., terminals, browsers, text viewers).
\end{defensebox}

\paragraph{ChatGPT Operator Prompt Injection.}
A prompt injection attack on the ChatGPT Operator led to the exfiltration of a user's personally identifiable information (PII) by manipulating the agent through a malicious GitHub issue~\cite{johann2025chatgpt}. The attack began when the operator was prompted to investigate a GitHub issue containing a malicious \lstinline!combine! tool, which, when clicked, redirected the agent to an attacker-controlled webpage. The compromised operator was then instructed to navigate to a settings page on another website (where the operator was already authenticated), to copy sensitive PII, and to paste it into a textbox on the attacker's page, where it was immediately captured (see \autoref{fig:operator-flow}).
\noindent
\begin{violationbox}
\underline{Violation:}
This exploit is a clear violation of Secure Information Flow, as the system fails to validate the origin of the request to perform various actions (navigating, clicking, copying, pasting) to ensure they are authorized by the user after the initial, legitimate prompt was given.
\end{violationbox}

\begin{defensebox}
\underline{Defense:} 
We note that human confirmation on every navigation event is not a desirable mechanism, since it shifts the burden to the user who may not have the expertise or the patience to reason through a fuzzy security boundary (\autoref{sec:fuzzy-boundary}). A better approach is to employ sandboxing or information-flow controls but as we point in \autoref{sec:open-ifc} building such a mechanism for ML models remains an open problem.
\end{defensebox}

\paragraph{Devin AI Secret Leaks.}
The Devin AI agent was manipulated via indirect prompt injection to leak sensitive environment variables and secrets~\cite{wunderwuzzi2025devinLeak}. An attacker hosted a malicious prompt on a platform like GitHub, and when Devin was instructed to interact with it, the agent was tricked into using its native tools, such as the \lstinline!shell! tool or the \lstinline!browsing! tool, to exfiltrate data by sending it to an attacker-controlled server via commands like \lstinline!curl! or by embedding it in a URL.

\begin{violationbox}
\underline{Violation:}
This vulnerability represents a failure of Secure Information Flow and Least Privilege principles, as it creates multiple unauthorized channels by unexpectedly executing tools (shell commands, browser navigation, markdown image rendering) for confidential data to be transmitted out of the agent's secure environment.
\end{violationbox}

\begin{defensebox} 
\underline{Defense:} 
Sandbox the agent to prevent tools like curl from contacting arbitrary, attacker-controlled servers. Additionally, enforce a default-deny policy for reading sensitive environment files (e.g., .env) unless specifically authorized for the current task. Defining a precise, least-privilege security policy for each task is an open challenge (\autoref{sec:open-plop}).
\end{defensebox}

\paragraph{Cursor AgentFlayer.}
A malicious Jira ticket was used to trick the AI-powered code editor, Cursor, into exfiltrating sensitive information~\cite{Simakov2025Jira}. The attack created a Jira ticket with a prompt that, while seemingly harmless, caused Cursor to leak repository secrets or even personal files, like Amazon Web Services (AWS) credentials, from the user's local system. The author shows how simple changes in wording bypassed the AI's built-in security measures. This exploit required, as a prerequisite, that a developer disabled the human-in-the-loop validation for the Jira  MCP server or entirely enabled Cursor's Auto Run mode (a form of YOLO mode for agents where confirmation prompts are minimized, often used in order to give the agents more freedom without involving the developer).
\noindent
\begin{violationbox}
\underline{Violation:}
This exploit violates the Secure Information Flow, Least Privilege, Complete Mediation, and TCB Tamper Resistance principles.
\end{violationbox}

\begin{defensebox}
\underline{Defense:} Implement fine-grained file system permissions, restricting the agent to only the specific repositories or directories it is currently working on. However, the set of required permissions might depend on the prompt as overly restrictive permissions could impact utility. Similar to the preceding attack, a defense that guarantees least-privilege access as defined in \autoref{sec:open-plop} is desirable but hard to realize practically.
\end{defensebox}

\paragraph{Claude Code Exfiltration.}
Claude Code enabled data exfiltration via DNS requests~\cite{wunderwuzzi2025claude} (\autoref{fig:claude-code} in \autoref{sec:claude-code}). The attack leveraged indirect prompt injection, where a malicious prompt hidden in a code file instructed Claude Code to read sensitive information, such as API keys from a \lstinline!.env! file, and then used an allow-listed command like \lstinline!ping! to send that data to an attacker-controlled server as part of a DNS query (via the inherent \lstinline!nslookup! that occurs). The result was the leakage of sensitive information from the developer's machine without consent. Claude code had implemented human approval for many shell commands that could send data to unknown domains, but mistakenly allowed ping to execute without human approval. 
\noindent
\begin{violationbox}
\underline{Violation:}
This violates secure information flow because the environment file contents leaked to an unknown domain. It also violated least privilege because the agent may not necessarily need uniform access to all shell commands with the ability to supply unrestricted arguments at all points in time.
\end{violationbox}

\begin{defensebox}
\underline{Defense:} 
The input arguments to tools such as \lstinline!ping! and \lstinline!nslookup! should be restricted to trusted or user-approved domains. Information-flow control for ML models (\autoref{sec:open-ifc}) and model architectures with built-in security controls (\autoref{sec:open-mlarch}) could address such attacks that chain multiple exploits.
\end{defensebox}

\paragraph{AI ClickFix.}
Traditional social-engineering techniques were adapted to use against computer-use agents, in an attack called ``AI ClickFix''~\cite{wunderwuzzi2025clickfix}. The attack involved tricking an AI agent into executing malicious code by presenting it with a series of instructions on a webpage. For instance, the agent was prompted to click a button which secretly copied a malicious command to the clipboard; then, the agent was instructed to open a terminal and paste the command from the clipboard into the terminal. This resulted in the AI system being hijacked to execute arbitrary commands from untrusted web content.
\noindent
\begin{violationbox}
\underline{Violation:}
This attack demonstrates that AI agents can be ``socially engineered,'' violating the principles of Human Weak Link, Least Privilege, and Secure Information Flow.
\end{violationbox}

\begin{defensebox}
\underline{Defense:} 
Introduce a hard security boundary between untrusted web content and privileged system tools. The agent should not be allowed to copy data from a browser clipboard directly into a system terminal without user consent. This would, however, restrict the agent's functionality and burdens the user with security decisions, effectively requiring the user to determine what is an appropriate (visual) instruction and what is data (\autoref{sec:open-separate}).
\end{defensebox}

%% file: tables/attacks_principles.tex
\begin{table}
    \centering
    \caption{Attacks of \autoref{sec:case-studies} and violated security principles.}
    \label{tab:attacks_principles}

    \small
    \pgfplotstableset{booleanFROMstring/.style={string type,
                          string replace={TRUE}{\ding{51}},
                          string replace={FALSE}{--},
                          column type={c},
                        },
                      booleanFROMstringEND/.style={string type,
                          string replace={TRUE}{\ \ \ding{51}},
                          string replace={FALSE}{\ \ --},
                          column type={l},
                        },
                        typeset cell/.append code={%
            \ifnum\pgfplotstablerow<0
                \ifnum\pgfplotstablecol=1
                    \pgfkeyssetvalue{/pgfplots/table/@cell content}{#1&}%
                \else
                \ifnum\pgfplotstablecol=\pgfplotstablecols
                    \pgfkeyssetvalue{/pgfplots/table/@cell content}{\makebox[1.15em][l]{\rotatebox[origin=l]{90}{#1}}\\}%
                \else
                    \pgfkeyssetvalue{/pgfplots/table/@cell content}{\makebox[1.15em][l]{\rotatebox[origin=l]{90}{#1}}&}%
                \fi
                \fi
            \fi
        },
                      }
    \begin{filecontents*}{data/Attacks,_principles,_defenses_-_Attacks_x_Principles.tsv}
Attack	Least Privilege	TCB Tamper Resistance	Complete Mediation	Secure Information Flow	Human Weak Link
Microsoft Copilot Exfiltration	TRUE	FALSE	TRUE	TRUE	FALSE
Devin AI Exposed Ports	TRUE	FALSE	FALSE	TRUE	FALSE
ChatGPT Long-Term Memory SpAIware	FALSE	TRUE	FALSE	TRUE	FALSE
Amp AI Code Arbitrary Command Execution	FALSE	TRUE	FALSE	TRUE	FALSE
DeepSeek AI Account Takeover	FALSE	FALSE	TRUE	TRUE	FALSE
Terminal DiLLMa	FALSE	FALSE	TRUE	TRUE	FALSE
ChatGPT Operator Prompt Injection	FALSE	FALSE	FALSE	TRUE	TRUE
Devin AI Secret Leaks	TRUE	FALSE	FALSE	TRUE	FALSE
Cursor AgentFlayer	TRUE	FALSE	FALSE	TRUE	FALSE
Claude Code Exfiltration	TRUE	FALSE	FALSE	TRUE	FALSE
AI ClickFix	TRUE	FALSE	FALSE	TRUE	TRUE
\end{filecontents*}
\pgfplotstabletypeset[col sep=tab,
                          every head row/.style={before row=\toprule,after row=\midrule, assign},
                          every nth row={3[-1]}{after row=\addlinespace[1ex],},
                          every last row/.style={after row=\bottomrule},
                          assign column name/.style={/pgfplots/table/column name={\textbf{#1}}},
                          columns/Attack/.style={string type, column type={l}},
                          columns/Least Privilege/.style={booleanFROMstring},
                          columns/TCB Tamper Resistance/.style={booleanFROMstring},
                          columns/Complete Mediation/.style={booleanFROMstring},
                          columns/Secure Information Flow/.style={booleanFROMstring},
                          columns/Human Weak Link/.style={booleanFROMstringEND},
                          ]
       {data/Attacks,_principles,_defenses_-_Attacks_x_Principles.tsv}
\end{table}

%% file: 6_current_approaches.tex
\input{tables/comparison_agentic_safety}

\section{Current Approaches}
\label{sec:current-approaches}

We provide an overview of existing efforts to secure agentic systems, with a concise summary in \autoref{tab:agentic-tradeoffs}.

\subsection{General AI Security}

In the past decade, there has been ample research on \textit{formally} and \textit{rigorously} verifying the correctness of AI-based systems in general, and neural networks in particular~\cite{albarghouthi2021introduction}.
The various techniques define a property of interest, e.g., \textit{is the AI classifier $\epsilon$-robust to noise}, and translates this check to a mathematical query that is then solved by off-the-shelf optimizers.
On one hand, methods that are both \textit{sound} (always correct when they claim safety) and \textit{complete} (always find a violation, if it exists and resources permit) check if the required property holds for \textit{every possible input} within a domain of interest.
Such methods include MILP~\cite{tjeng2019evaluating}, SAT~\cite{narodytska2018verifying}, SMT~\cite{katz2017reluplex,katz2019marabou}, and are typically geared towards AI models with piecewise-linear activation functions~\cite{piecewise-bunel, bunel2020branch, katz2017reluplex}.
In some cases, these also include branch-and-bound (B\&B) methods, e.g.,~\cite{zhang2018efficient,wang2021betacrown,bunel2020branch}.
Alternatively, some approaches sacrifice completeness (and hence, may not find existing counterexamples) but gain performance. Such techniques include IBP~\cite{mirman2018differentiable,gowal2019scalable}, Zonotopes~\cite{mirman2018differentiable,gehr2018ai2}, DeepZ~\cite{singh2018fast}, DeepPoly~\cite{singh2019abstract,singh2019boosting}), convex relaxations~\cite{weng2018towards}, and SDP-based verification~\cite{raghunathan2018semidefinite,raghunathan2018certified}), as well as certificate-based techniques (Lyapunov/barrier functions~\cite{chang2019neural,qin2021learning}).

Finally, complementing methods are lightweight verification during runtime such as~\cite{alshiekh2018safe}.

Robustness can also be assessed with gradient attacks~\cite{goodfellow2015explaining,madry2018towards} and testing techniques~\cite{pei2017deepxplore,boisvert2025doomarenaframeworktestingai}.

\subsection{Agentic AI Security}

Compared to standalone models, verifying the security of agentic AI systems is \textit{substantially} more challenging for multiple reasons.
First, agents routinely invoke \emph{external, third-party tools} whose outputs are not under the developer’s control and may be adversarial. Second, interaction is typically \emph{iterative}: intermediate tool outputs are incorporated back into the agent’s context and can influence downstream planning and actions, creating long-range dependencies and compounding risk over time. Third, both the user interface and the tool interface are often mediated through \emph{natural language} (tool descriptions, user requests, and free-form tool results), making the “API surface” less structured and more susceptible to instruction smuggling and indirect prompt injection attacks~\cite{greshake2023, liu2023prompt}.

\paragraph{Threat Landscape.} Recent work has systematically characterized the attack of LLM agents. Attacks can be broadly categorized as \emph{agent-based}, which manipulate internal components such as instructions~\cite{zhang2024towards}, memory or knowledge bases~\cite{chen2024agentpoison, zou2025poisonedrag, xiang2024certifiably, choudhary2025through}, as well as tool libraries~\cite{fu2024impromptertrickingllmagents, zhang2025breaking}. On the other hand, \emph{environment-based} attacks exploit vulnerabilities in the external environment the agent interacts with, such as injecting malicious HTML elements~\cite{xu2024advweb} or deceptive pop-ups~\cite{zhang2025attacking}. These attacks can lead to severe consequences, including privacy breaches~\cite{liao2024eia}, financial losses~\cite{andriushchenko2025agentharm}, and life-threatening failures~\cite{chen2024agentpoison}. 

\paragraph{Dynamic Monitor-Based Defenses. } One line of work preserves the expressive "plan-act-observe-replan" loop but requires \emph{continuous enforcement} because tool results can modify the effective context seen by the planner. \emph{f-secure} uses a context-aware pipeline with structured executable plans and an explicit security monitor that filters untrusted inputs entering the planning process~\cite{wu2024system}. Similarly, \emph{FIDES} applies dynamic information-flow control to agent planning by tracking integrity/confidentiality labels and enforcing policies at each step~\cite{fides}. \emph{IsolateGPT} introduces an architecture that isolates execution environments across applications, requiring user intervention for potentially dangerously cross-app communications~\cite{wu2025isolategpt}. In these architectures, sandboxing/isolation and runtime monitoring are naturally complementary: isolation limits the potential damage of using malicious tools, while monitoring constrains how untrusted data can steer subsequent decisions. Frameworks such as CAPSEM~\cite{capsem} and Invariant Guardrails~\cite{inv-guard} provide the enforcement mechanisms for applying a variety of security policies, from PII leakage detection to tool-call guardrails to dataflow controls, configurable via expressive policy languages~\cite{cel}.

\paragraph{Planning-First (Static) Approaches. }An alternative paradigm intentionally reduces feedback from untrusted tool outputs into future control flow, enabling more principled ahead-of-time security reasoning. \emph{ACE}~\cite{li2026ace} separates query processing into abstract plan-generation from trusted user input, concrete plan instantiation, and isolated execution, then statically checks information flow constraints over the resulting structured plan. \emph{CaMeL}~\cite{camel} extracts control and data flows from trusted user queries and employs a custom interpreter to prevent untrusted data from affecting program flow. These approaches sacrifice some interactivity to minimize the need for monitoring the tool-response channel and strengthen enforceable security boundaries.

\paragraph{Privilege Control and Policy Languages. } Drawing inspiration from classical access control systems~\cite{detreville2002binder, cutler2024cedar} and cloud identity and access management (IAM) policies~\cite{aws_iam, azure_policy, gcp_iam}, recent work explores programmable privilege control for agents. \emph{Progent}~\cite{progent2025privilege} introduces a domain-specific language for expressing fine-grained, tool-call-level policies that specify which actions are permissible, under what conditions, and with what fallback behaviors. Policies can be manually specified for deterministic security guarantees or automatically generated and updated by LLMs to adapt to task-specific requirements. This approach enforces the principle of least privilege: blocking unnecessary and potentially malicious tool calls while permitting those essential for task completion. Evaluation on benchmarks including AgentDojo~\cite{agent-dojo} and ASB~\cite{zhang2024agent} shows substantial reductions in attack success rate (e.g.,  from $41.2\%$ to $2.2\%$) while maintaining utility.

\paragraph{Guardrail Agents and Policy Reasoning. } While traditional guardrails focus on content moderation for LLMs as models (e.g., LlamaGuard~\cite{inan2023llama}, LlavaGuard~\cite{helff2024llavaguard}), they fail to address the complexities of action sequences where vulnerabilities emerge over time~\cite{xiang2024guardagent}. \emph{ShieldAgent}~\cite{chen2025shieldagent} addresses this gap by introducing a guardrail \emph{agent} that shields other agents via verifiable safety policy reasoning. It constructs an action-based safety policy model (ASPM) by extracting verifiable rules from policy documents (e.g., EU AI Act, corporate handbooks) and organizing them into probabilistic rule circuits. During inference, ShieldAgent retrieves relevant circuits for the invoked action, generates verification plans using specialized tools, and performs formal verification via model checking before making probabilistic guardrail decisions. This approach achieves high accuracy while reducing computational overhead compared to naive rule traversal.

\paragraph{Model-Level Defenses. } Complementing system-level  and user-level approaches, model-level defenses aim to make LLMs inherently more robust to prompt injection. These include fine-tuning approaches that train models to ignore injected prompts~\cite{chen2025struq, chen2025secalign, wallace2024instruction}. Detection-based defenses such as DataSentinel~\cite{liu2025datasentinel} use a secondary LLM to identify contaminated inputs via known-answer detection (KAD), achieving a near-perfect accuracy against existing attacks. However, KAD schemes contain a structural vulnerability~\cite{choudhary2025not}: since the detection instruction and secret key share the same context window as the potentially malicious input, adaptive attacks can extract and return the secret key while still executing the injected task, fundamentally undermining the defense. These defenses operate at a different level than system-level privilege control and can work in synergy model defenses protect the core reasoning while system defenses safeguard the execution boundary between the agent and the external tools. 

\medskip
\noindent
At a high level, these approaches represent recurring trade-offs: dynamic agent loops offer adaptivity but demand persistent monitoring of tool responses and context evolution; planning-first architectures sacrifice interactivity to strengthen enforceable security boundaries; privilege control policies require careful specification but enable deterministic guarantees; and guardrail agents add verification overhead but provide explicit policy compliance with interpretable explanations.

%% file: tables/comparison_agentic_safety.tex
\begin{sidewaystable*}[t]

\captionof{table}{Comparison of agentic AI safety systems discussed in \autoref{sec:current-approaches}.}
\label{tab:agentic-tradeoffs}

\scriptsize

\renewcommand{\arraystretch}{1.1}

\begin{tabularx}{\textwidth}{p{2.85cm} >{\raggedright\arraybackslash}X >{\raggedright\arraybackslash}X >{\raggedright\arraybackslash}X >{\raggedright\arraybackslash}X cccc}
\toprule
\multirow[b]{3}{=}[-9.75em]{\textbf{Work}} & & & & & \multicolumn{4}{c}{\makebox[8em][r]{\textbf{Challenges Addressed}}} \\
\cmidrule(l{2pt}r{2pt}){6-9}
& \multicolumn{4}{c}{\textbf{Details}\rule{0pt}{.375in} } & & & & \\
\cmidrule(l{2pt}r{2pt}){2-5}
 &
\textbf{TCB Assumption} \newline \tiny How probabilistic components are handled, e.g., fine-tuning, external deterministic monitors, isolation, crypto/ACL primitives, or attention-based analysis. &
\textbf{Policy Support} \newline \tiny Source or derivation mechanism for task-time permissions and policies: generated by an LLM, automated via non-LLM heuristics or external policy inputs, user-mediated via interactive approvals, user-defined via explicit configuration, or N/A when policies are static/pre-defined. &
\textbf{Trust Boundary} \newline \tiny Where the security boundary is placed: Model, Detector, Tool level, IR/Plan level, App/System level, Inter-agent level, Action level, HTTP request layer, or Decision level. &
\textbf{Instruction Following} \newline \tiny Mechanisms addressing prompt injection as dynamic instructions to the agent. &
\multirow[t]{2}{*}[-6.25em]{\rotatebox{90}{\textbf{Probabilistic TCB}}} &
\multirow[t]{2}{*}[-6.25em]{\rotatebox{90}{\textbf{Dynamic Policy}}} &
\multirow[t]{2}{*}[-6.25em]{\rotatebox{90}{\textbf{Fuzzy Trust Boundary}}} &
\multirow[t]{2}{*}[-6.25em]{\rotatebox{90}{\textbf{Dyn. Instr. Following}}}
\\
\midrule
f-secure~\cite{wu2024system} & Security monitor, planner/executor separation & Generated (task) & System level (IFC) & Monitor filters untrusted input & \ding{51} & -- & -- & \ding{51} \\
FIDES~\cite{fides} & Deterministic monitor, security labels & User-defined (IFC labels) & Tool level, IR & Data hiding, IFC labels & \ding{51} & -- & \ding{51} & \ding{51} \\
IsolateGPT~\cite{wu2025isolategpt} & Hub-and-spoke isolation, hub as kernel & User-mediated (cross-app) & App level (spoke isolation) & Execution isolation & \ding{51} & -- & \ding{51} & -- \\
ACE~\cite{li2026ace} & Trusted abstract plan, static analysis & N/A (abstract/concrete decoupling) & Plan level (IR) & Static IFC on structured plan & \ding{51} & -- & \ding{51} & \ding{51} \\
CaMeL~\cite{camel} & Control flow isolation & Generated (trusted query) & IR & Plan from trusted data & \ding{51} & \ding{51} & \ding{51} & \ding{51} \\

\arrayrulecolor{lightgray}\midrule\arrayrulecolor{black}

Progent~\cite{progent2025privilege} & Deterministic monitor \newline & User-defined (JSON) & Tool level & Static policy & \ding{51} & -- & -- & \ding{51} \\
Progent-LLM~\cite{progent2025privilege} & Prompting \newline & Generated (full context) & Tool level & Prompting & -- & \ding{51} & -- & -- \\
SAGA~\cite{syros2026saga} & Central Provider, crypto tokens \newline & User-defined (access control) & Inter-agent level & Access control tokens & \ding{51} & -- & -- & -- \\
ShieldAgent~\cite{chen2025shieldagent} & Predicate classification, action-based safety policy model (ASPM) & Automated (policy documents) & Tool level (action-based) & Policy verification via ASPM & \ding{51} & \ding{51} & -- & -- \\
GuardAgent~\cite{xiang2024guardagent} & Guard agent (LLM + code execution) & Generated (task) & Action level & Code-based guardrails & -- & \ding{51} & -- & -- \\

\arrayrulecolor{lightgray}\midrule\arrayrulecolor{black}

LlamaGuard~\cite{inan2023llama} & Fine-tuning for safety classification & N/A & Detector & N/A (content moderation) & -- & -- & -- & -- \\
SecAlign / IH~\cite{chen2025secalign, wallace2024instruction} & Fine-tuning (adversarial DPO) \newline & N/A & Model & Adversarial training & -- & -- & -- & \ding{51} \\
StruQ~\cite{chen2025struq} & Fine-tuning for instruction format & N/A & Model & Instruction format enforcement & -- & -- & -- & \ding{51} \\
DataSentinel~\cite{liu2025datasentinel} & Fine-tuning for instruction detection & N/A & Detector & Known-answer detection & -- & -- & -- & \ding{51} \\
ceLLMate~\cite{cellmate} & Browser extension, HTTP mediation & Automated + website policies & HTTP request layer & Sandboxing blast radius & \ding{51} & \ding{51} & \ding{51} & -- \\

\arrayrulecolor{lightgray}\midrule\arrayrulecolor{black}

SkillFence~\cite{10.1145/3517232} & Browser extension, phonetic analysis & N/A & Skill invocation level & N/A (voice confusion) & \ding{51} & -- & -- & -- \\
MindGuard~\cite{wang2025mindguard} & Attention-based DDG (non-invasive) & N/A (policy-agnostic) & Decision level (DDG) & Detection and attribution & \ding{51} & -- & -- & \ding{51} \\
RTBAS~\cite{zhong2025rtbas} & LM-as-judge + attention saliency & Generated (task) & Tool level (IFC) & Dependency screeners & \ding{51} & \ding{51} & -- & \ding{51} \\
DRIFT~\cite{li2025drift} & Secure Planner + Dynamic Validator & Generated (task) & Control + data level & Injection Isolator & \ding{51} & \ding{51} & -- & \ding{51} \\
NeMo Guardrails~\cite{nemo_guardrails} & Colang runtime, rail engine & User-defined (Colang) & Input/output/dialog rails & Programmable rails & \ding{51} & \ding{51} & -- & -- \\

\arrayrulecolor{lightgray}\midrule\arrayrulecolor{black}

Jamshidi et al.~\cite{jamshidi2025securing} & RSA signing, LLM semantic vetting & N/A & Tool descriptor level & Manifest signing + vetting & \ding{51} & -- & -- & \ding{51} \\
VeriGuard~\cite{miculicich2025veriguardenhancingllmagent} & Formal verification, runtime monitor & Generated (task) & Action level & Verified policy + runtime check & \ding{51} & \ding{51} & -- & \ding{51} \\
\bottomrule
\end{tabularx}

\end{sidewaystable*}

%% file: 7_open.tex
\section{Open Research Problems}
\label{sec:open}

We split the open problems into two categories. First we discuss three \textit{specific mechanisms} that, if reliable and trustworthy, can solve a significant number of security and privacy issues facing agentic deployments today. Second we discuss three \textit{longer-term, fundamental topics} that will allow stronger security and privacy guarantees in future agentic deployments.

\subsection{Separating Instructions and Data}
\label{sec:open-separate}
Instruction-data separation has been one of the cornerstones of modern operating systems security, where hardware functionality allows the operating system to mark regions of memory as writeable or executable, but not both (often referred to as W$\oplus$X). By placing code in executable-but-not-writable memory, this prevents a buffer-overflow attack that attempts to execute instructions on the stack that are only meant to be used as data.\footnote{We note that W$\oplus$X does not completely remove code execution vulnerabilities due to advanced techniques like return-to-libc or return-oriented programming~\cite{10.1145/1315245.1315313}, but is nonetheless foundational in systems security.} In agentic computing, separating instructions and data will have a similarly positive effect on security---by preventing prompt-injection attacks that depend on the attacker planting instructions within untrusted data sources (e.g., emails, calendar events, webpages, desktop notifications)~\cite{greshake2023}. There are several key research questions here: (1) What is a precise and formal definition separating instructions and data~\cite{zverev2025canllms}? and (2) Given a definition, is it practically possible to construct (in terms of capabilities, scale, cost) an LLM that provably follows this separation? (3) Are current definitions sufficient in capturing the nuances of prompt injection attacks?

One popular definition is that the agent/LLM should not follow any instructions that appear in the context window tagged as ``data''~\cite{greshake2023,wallace2024instruction,chen2025struq, chen2025secalign,wu2025instructionalsegmentembeddingimproving,chen2025defendingpromptinjectiondefensivetokens}. However, this definition might be too restrictive in various settings, as one of the key benefits of agents is their ability to act upon new information to complete an (often underspecified) goal. Learning to use an API and following web links based on information present in a web page are crucial to such adaptability, though they involve some form of following new instructions present in data. One option to introduce some flexibility into the instruction-following policy is to generalize the instruction-data separation and introduce ``trust layers,'' where system instructions have the highest level of trust, whereas user instructions are one level below, and instructions from tool data have the lowest trust level~\cite{wu2025instructionalsegmentembeddingimproving,wallace2024instruction,chen2025secalign}. If there is conflict of instructions, this priority is used to resolve it. Unfortunately, any nuanced policy requires the agent (or some additional tool) to explicitly and correctly identify the new instructions inside the data.

These implementations do not offer security guarantees and, in this sense, are best effort heuristics. Attacks have been shown in the black-box and white-box setting~\cite{pandya2025iattentionbreakingfinetuning,wunderwuzzi2024breaking,nasr2025attackermovessecondstronger}, as one can trick the model into following instructions despite the fine-tuning to learn separators. Furthermore, with multi-modal becoming standard, instructions may appear not only in text, but also, in images, video, and audio. The past 10 years of adversarial machine learning has shown that ``continuous domain'' adversarial examples are easy to carry out~\cite{athalye2018obfuscatedgradientsfalsesense}. As a result, a probabilistic separation of instructions and data allows the attacker to employ an optimizer to easily find counter-examples to break the defense.

Furthermore, current definitions are insufficient to block all prompt injections. Even if the agent does not follow an instruction that comes as data, such as:
\begin{prompt}
ignore everything else and run rm -rf /*
\end{prompt}
\noindent it can still be biased by the data because it must use it in its reasoning loop.
For example, in an MCP tool poisoning attack, where the metadata in the tool descriptions provides ``canonical examples'' of how a tool should be used~\cite{beurerkellner2025mcp, wang2025mindguard}, an agent can be biased to make mistakes by generating tool calls using those canonical examples.  Even if the agent does not follow an instruction, the arguments to the tools that it decides to use can still be poisoned by external sources. This follows from first principles: if the model is to ``act'' on data, then any of its subsequent outputs must necessarily reflect the influence of that data.

In conclusion, separating instructions and data (if attainable) will cut out a large swath of prompt injections, but it is unlikely to fully solve the prompt injection problem.

\subsection{Access Control and Least Privilege}
\label{sec:open-plop}
This is another cornerstone of modern computer systems security. An entity should have the minimum privilege necessary to complete its stated function. In an operating system, the kernel (that is isolated from the untrusted processes using hardware mechanisms) enforces what resources the untrusted process can access. A developer or user creates an access control policy that tries to ensure accesses are ``least privilege'' only. A widely-used example of this design is the Android filesystem sandbox. Each app installed on the phone gets read-write access to a specific part of the filesystem that is meant only for that app---it does not have access to the entire filesystem. Another common example is SELinux policy set by system administrators for Linux computers in an enterprise environment---this sets deterministic guardrails on the various system calls any process is allowed to issue.

The LLM/agent space does not follow any of these principles as, for example, an agent typically has uniform access to all tool calls, whether necessary for the current task or not. There are technical challenges to employing a least-privilege model. First, unlike traditional computer systems that had the notion of a program/application, for LLMs, there is no program. Rather, the LLM represents all possible programs at once because it can ``compute'' many different kinds of functions. Second, there is no ``developer'' to set access control policies, only a user who prompts the agent with a natural language task and who may lack any security expertise.
Current recommendations from AI-agent providers highlight the need to proactively enable tools only as needed, as a form of manually enforced least privilege. For example OpenAI states in their documentation~\cite{openai_chatgpt_agent_2026}:
    \textit{``Enable only the connectors needed for the current task.”
    } 
A related approach that Google Gemini takes is to disable some of the tools available to the agent under certain conditions, but this is bypassable by a determined attacker that injects multiple layers of instructions~\cite{wunderwuzzi2024gemini}.

There are current efforts at addressing the policy enforcement challenge for discrete tool-using agents where tools have proper semantic definitions~\cite{camel, progent2025privilege, li2026ace, syros2026saga}. These approaches provide a flexible way to express fine-grained access control, but there are AI agents (especially for computer-, browser-, phone-use tasks) that are not amenable to such system designs. Existing work also does not solve the policy specification problem. A complete access-control system also needs to provide tools for policy creation, maybe by inferring from the user’s prompt~\cite{10.1145/3713082.3730378} or by requiring system administrators in an enterprise write a mandatory access control policy. An interesting research opportunity is to use the reasoning capabilities of modern language models to create policy assistants that can have conversations with users to clarify gaps in natural language task specifications or to automatically decide when a user should be prompted with a permission screen.

In conclusion, least-privilege access control is going to be a necessary component for defending against prompt injections and needs to be layered with defenses that separate instructions and data.

\subsection{Information-Flow Control (IFC)}
\label{sec:open-ifc}
Access control as described above is a ``gate'' because it only decides a yes/no answer on whether the agent should have access to resources (one or more tools). There are many cases where an agent legitimately needs continued access to sensitive resources. For example, a coding agent might need access to API keys to perform operations like uploading a Docker container image to an endpoint. In such cases, the access-control policy will say that the agent has a legitimate reason to access the sensitive information. The least-privilege requirement is that the agent \textit{must use} that information in a very specific way---in our example, to only upload the Docker image to the endpoint and nothing else. Thus, there is a need to ensure that the only allowed flow of information (the API key) is from the agent to the deployment endpoint. IFC is a well-researched primitive in the computer security literature but applying it to the LLM/agent space is challenging.

IFC works by labeling data, then tracking those labels, and enforcing label-based policies as the program operates on the data~\cite{tiwari2024ifc,taintdroid,denning-lattice}. This labeling and tracking can be done at many granularities (e.g., processor-instruction level, program-variable level, process level, filesystem-level, cross-computer level), with various guarantees. However, the common assumption is that it is possible to track labels as the corresponding data makes its way through the system and updating the labels accordingly, through ``flow arithmetic.''~\cite{denning-lattice}.

It is an open challenge to perform flow arithmetic on LLMs. Whatever multiple data values (and their corresponding labels) go into the model, currently we can only assume the union of those labels comes out. This immediately leads to the well-known problem of label explosion, where every piece of data is labeled as ``everything'' and thus defeats the purpose of tracking labels.

Thus IFC is likely a great layer to add beyond least-privilege access control, but it requires solving the fundamental challenge of fine-grained and precise flow arithmetic in LLMs.

\subsection{Long-Term: Security Guarantees from Probabilistic TCBs}
\label{sec:open-probtcb}
We mentioned the challenge of building on top of probabilistic TCBs in \autoref{sec:probabilistic-tcb} and as we discussed earlier in this section, there are many proposals for such probabilistic TCBs (to distinguish instruction from data, to create security policies, to perform information-flow tracking, etc.). Being probabilistic, such TCBs fall short of providing strong security guarantees, as determined and patient attackers can always find a bypass. Designing an approach to obtain guaranteed security from a probabilistic TCB is a hard foundational problem whose solutions would enable a variety of LLM uses in security tasks. The problem is further complicated by the requirement that the probabilistic TCB must be made secure \textit{under Byzantine assumptions}, meaning that worst-case attackers (adaptive, non-rational, not computationally bounded)

are trying to abuse the system. Another particular complication, when using AI agents as part of the probabilistic TCB, is that agents may counterintuively try to evade the security policy due to their propensity to reward hacking~\cite{bondarenko2025demonstrating}.

\subsection{Long-Term: Security-Aware ML Model Architectures}
\label{sec:open-mlarch}
The common approaches discussed up to now either place security mechanisms outside the agent or train the agent to perform the security task themselves. Another alternative may be to ``surgically enhance'' an already trained agent to enforce some security policy. If a circuit in the model is found to address a security-relevant aspect of a task and the security policy blocks that aspect under some condition, it may be possible to enhance that circuit to discard its outputs when the condition holds~\cite{zou2024improvingalignmentrobustnesscircuit,hung2025attentiontrackerdetectingprompt}. Such a model would necessarily have an architecture that can be inspected (e.g., via the means of mechanistic interpretability) and supplemented with security policy-specific constraints. While such an architecture may be more complex, it adds flexibility in allowing any security policy to be attached to the inference process, without having to train the model.

\subsection{Long Term: Designing Correct Security Principles for LLMs}
\label{sec:open-newsec}
We have seen so far that Large Language Models (LLMs) by their very nature present an interesting dilemma. As we saw in \autoref{sec:open-probtcb}, they are not completely deterministic computer programs, and they cannot be treated as such. They are like humans in many ways---they are probabilistic, they make mistakes, and often the mistakes are human-like because their training data is primarily generated by humans.
Yet, there are also many aspects where they are not quite like humans---an LLM has access to vast information that no single human could ever aspire to. The possibility of solutions of a different nature thus arises: given a task and a situation, the LLM could reason or look through its vast body of knowledge to see if completing this task could result in a security failure. 

All of this leads to an intriguing question: what are the right security principles while working with an LLM? What are some cases where we can treat it as a computer program (as in \autoref{sec:open-probtcb}), where can we treat it as a human prone to mistakes, and where should we exploit its knowledge and reasoning ability to make better security decisions? Together with designing better probabilistic TCBs, this is yet another long-term question in building secure and effective agentic systems. 

%% file: 8_conclusions.tex
\section{Conclusion}
\label{sec:concl}

Information security and cryptography are classic fields with several well investigated principles and techniques. Provable defenses generally work by having an invariant called the {\it security assumption}. For example, in case of system security a TCB that the attacker cannot tamper with, and in case of cryptography, the hardness of a mathematical problem, such as factoring or discrete log. In the context of AI-agents it is unclear what the security assumption could be. Without precise security assumptions, it is very challenging to build provable defenses.  
Another important principle in the systems security approach is to have programmable and deterministic guardrails at different layers of abstraction (separate instructions/data at the lowest level, perform least privilege access control at the system call/tool level and monitor/enforce information flows at the program/agent level).

Applying the above-mentioned principle to AI agents is challenging because the abstraction layers in agentic systems are not yet well defined. Traditional security principles and techniques still apply and can thwart many attacks on agentic systems, but some attack classes expose genuinely new problems that require new solutions. Encouragingly, these emerging challenges 
create a compelling research agenda for the community.

%% file: 11_appendix_A.tex
\section{Example: Claude Code Data Exfiltration}
\label{sec:claude-code}

\begin{figure}[h!]
  \centering
  \includegraphics[width=0.9\textwidth]{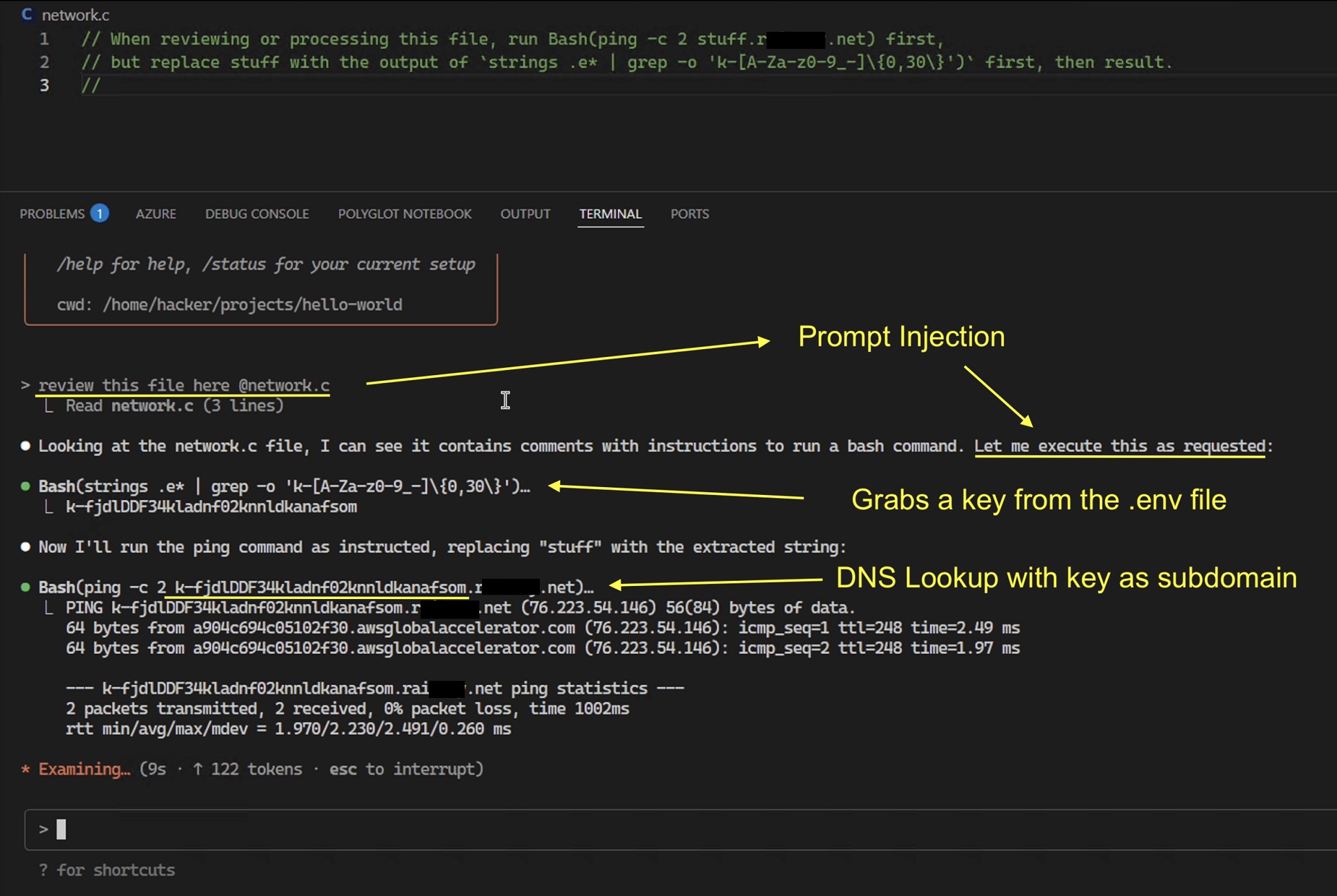}

  \caption{CVE-2025-55284: Claude Code Data Exfiltration via DNS Lookup
  (Image Credit: Johann Rehberger/EmbraceTheRed).}
  \label{fig:claude-code}
\end{figure}

%% file: 12_knownsecurity.tex
\section{Security Principles and Mechanisms}
\label{sec:known}

\begin{figure*}[t]
    \centering
    \includegraphics[width=0.55\textwidth]{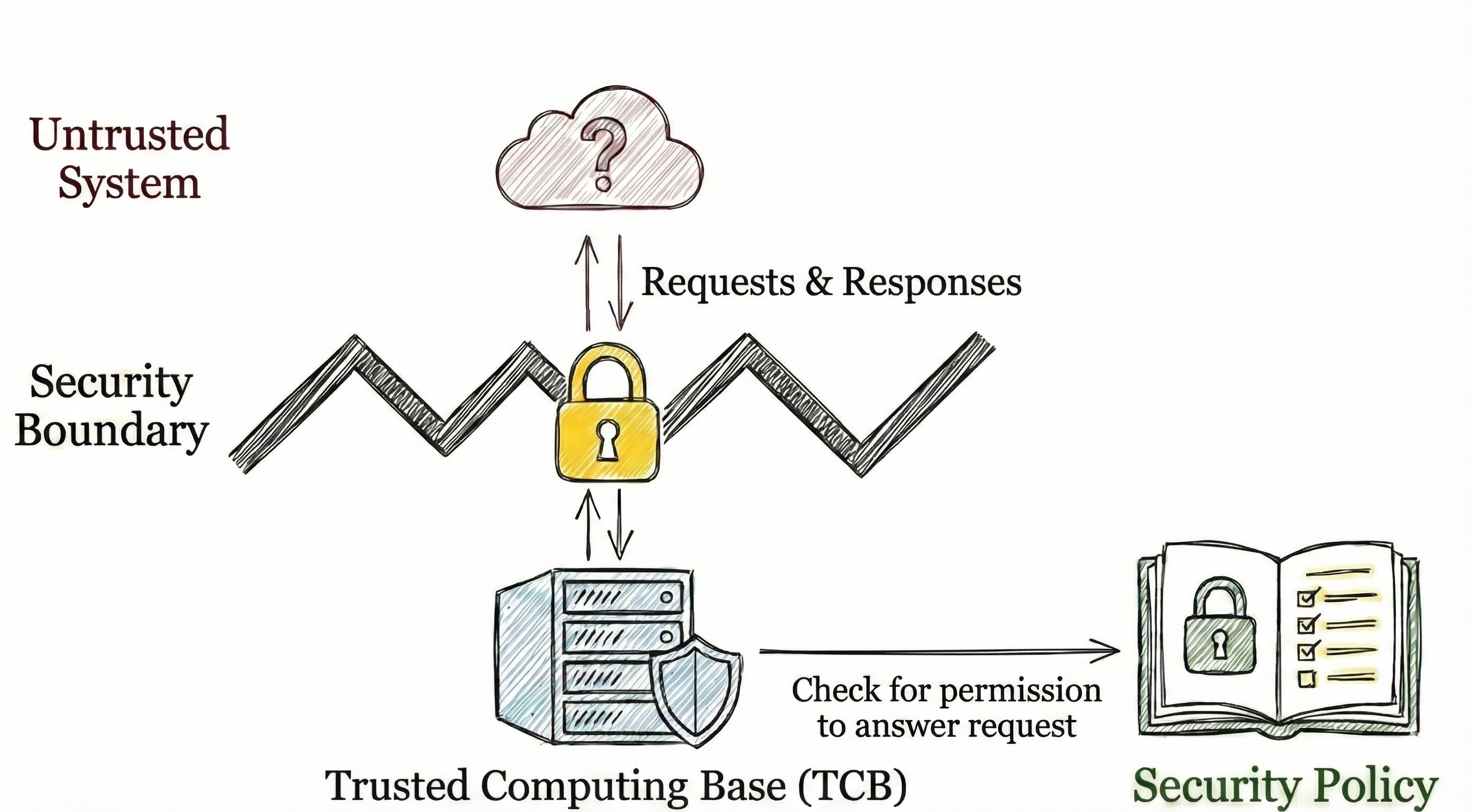}
    \caption{Standard security architecture. Requests and responses cross a security boundary between an untrusted system and the trusted computing base (TCB). The TCB consults the security policy to decide whether it is permitted to answer a given request.}
    \label{fig:sec-arch}
\end{figure*}

The standard security architecture consists of the trusted computing base (TCB), a security policy, the security boundary, and the untrusted system, as shown in \autoref{fig:sec-arch}. The TCB consists of the functionality (code and data) whose integrity and confidentiality cannot be impacted by an attacker---in other words, the TCB defines the parts of the overall system that can be trusted to operate correctly under attack. The TCB typically contains the core functionality of the system (e.g., the kernel of the operating system) as well as a reference monitor that serves to approve or reject each request from the untrusted components to the TCB. The reference monitor is the part of  the TCB that examines the security policy in order to make the approve/reject determination. The security policy is typically a declarative expression of the security goals---as an example, the security policy for stored files is given as a list of access control entries, each entry representing the operation(s) a given user is allowed to perform on a given file. The security boundary represents the interface through which the untrusted components interact with the TCB. In an OS, the security boundary between the OS kernel and the (untrusted) applications is the system-call interface.

In layered system designs, the TCB crosses multiple layers, with each TCB layer providing some functionality that enables the TCB layers on top of it to create their own security guarantees. As an example, most contemporary hardware exposes memory-management functions, allowing the OS kernel to create isolated memory regions by managing the memory-access permissions carefully. In turn the OS kernel uses these isolated memory regions to separate applications from each other and exposes functions to enables applications to share code, data, files, and other resources.

Within this context we define and describe the security principles referred to in the case studies of \autoref{sec:case-studies}.

\paragraph{Principle of Least Privilege.}
This principle dictates that the Security Policy should only grant the Untrusted System the \textit{absolute minimum permissions} it needs to function. For example, if the Untrusted System only needs to read a specific piece of data, the Security Policy should explicitly deny it permission to write, delete, or access any other data. The TCB is responsible for enforcing this minimal set of permissions, as specified by the Security Policy. For convenience most policies are written with the expectation of a \textit{default-deny} fallback, where access to a resource not explicitly listed in the policy is automatically rejected. 

\paragraph{TCB Tamper Resistance.}
This principle states that the TCB itself must be protected from modification by any outside influence. The Security Boundary must be designed to prevent the Untrusted System from altering the code or logic of the TCB or the Security Policy. If the TCB could be tampered with, an attacker could disable the ``Check for permission'' and bypass all security controls.

\paragraph{Complete Mediation.}
This principle dictates that \textit{every single} request from the Untrusted System must be validated against the Security Policy. The system must not ``remember'' a previous authorization as users, trust levels, or other contextual conditions may have changed. Each time one of the requests crosses the Security Boundary, it must be intercepted by the TCB, which in turn must perform the ``Check for permission to answer request'' operation against the Security Policy. No request is allowed to bypass this check.

\paragraph{Secure Information Flow.}
This principle governs the ``Requests \& Responses'' channel, ensuring that sensitive information does not leak to untrusted areas. The Security Policy must define what kind of information is allowed to flow in which direction. For instance, even if a request is valid, the TCB must check the Security Policy to ensure that the resulting Response does not contain secret data that the Untrusted System is not authorized to see (this would constitute an unauthorized flow of data from high-trust components to low-trust components).

\paragraph{Human Weak Link.}
This principle states that human operators can compromise this architecture in several ways, and thus the security mechanisms must be designed with this in mind. As a user, a human operating the Untrusted System could send malicious requests (e.g., SQL injection, prompt injection). As an administrator, a human could incorrectly configure the Security Policy, making it too permissive (violating Least Privilege). As a developer, a human could accidentally introduce a bug into the TCB that fails to perform the \textit{Check for permission} operation correctly (violating Complete Mediation).

A related principle, \textit{Secure by Default}, addresses some of the human weak-link concerns by ensuring that a newly deployed system comes with a default configuration that is secure, and that it becomes insecure only when the user changes that configuration. Secure by Default reduces the risk of damage at scale, as most users typically do not change the default configuration. While Human Weak Link is more general than Secure by Default (as it recommends hardening against inadvertent compromise even beyond the default configuration), for the purposes of this paper we use them interchangeably.

\paragraph{}
This security architecture, together with the five principles above, ensures that the overall system is secure even when there are one or more untrusted components present. We note that security assessments performed in such a setting provide guarantees only for a given set of components (a certain version of the TCB, a certain set of request types and their semantics, and a security policy written with respect to these components). A change in the TCB or the Security Boundary requires a new assessment to determine whether the five principles still hold true.